\documentclass[12pt]{article}

\usepackage{epsfig}
\usepackage{scicite}

\def\SrCu{Sr$_{14}$Cu$_{24}$O$_{41}$\,}

\def\SrCax{Sr$_{14-x}$Ca$_{x}$Cu$_{24}$O$_{41}$\,}
\def\cm-1{cm$^{-1}$}

\newenvironment{sciabstract}{%
\begin{quote} \bf}
{\end{quote}}

\newcounter{lastnote}

\title{Sliding Density-Wave in \SrCu Ladder Compounds}

\author
{G.~Blumberg,$^{1\ast}$ P.~Littlewood,$^{1,2}$ A.~Gozar,$^{1}$
B.S.~Dennis,$^{1}$\\ 
N. Motoyama,$^{3}$ H. Eisaki,$^{4}$ and S.~Uchida$^{3}$\\
\\
\normalsize{$^{1}$Bell Laboratories, Lucent Technologies, Murray Hill, NJ 07974,
USA}\\
\normalsize{$^{2}$University of Cambridge, Cavendish Laboratory, Cambridge, 
CB3 0HE UK}\\
\normalsize{$^{3}$The University of Tokyo, Bunkyo-ku, Tokyo 113, Japan}\\
\normalsize{$^{4}$Stanford University, Stanford, CA 94305, USA}\\
\\
\normalsize{$^\ast$To whom correspondence should be addressed; E-mail: 
girsh@bell-labs.com.}
}

\date{ }

\begin{document} 


\baselineskip24pt


\maketitle


\newpage 
\begin{sciabstract}
We used transport and Raman scattering measurements to identify
the insulating state of self-doped spin $\frac{1}{2}$ two-leg ladders of 
\SrCu as a weakly pinned, sliding density wave with non-linear conductivity 
and a giant dielectric response that persists to remarkably high 
temperatures.  
\end{sciabstract}


Investigation of the charge and spin dynamics of spin 
$\frac{1}{2}$ low-dimensional
transition metal oxide materials is attracting attention because
of the critical nature of their ground state and the relevance to the phase
diagram of high temperature superconducting cuprates
\cite{Rice96,Dagotto99}.
The competition between insulating states at low hole concentrations 
and superconductive pairing at higher hole densities 
has emerged as a key feature of the high-T$_{c}$ (critical 
temperature) problem, but the character 
of the insulating states has remained elusive.
Electronically one-dimensional (1D) materials are susceptible to a drastic
change of their ground state to an insulator with spontaneous broken symmetry. 
For example, Peierls ordered states in which the links connecting 
nearest neighbor sites acquire modulated values for their charge and/or 
spin densities as well as for the exchange coupling constants are associated 
with broken translational symmetries and have been extensively discussed in 
the literature \cite{Sachdev}.
In the charge ordered state, collective excitations --- such 
as those seen previously in sliding charge- and spin-density wave (C/SDW) 
compounds --- should dominate the low energy dynamics.

\begin{figure}[ht]
    \baselineskip24pt
\centerline{
\epsfig{figure=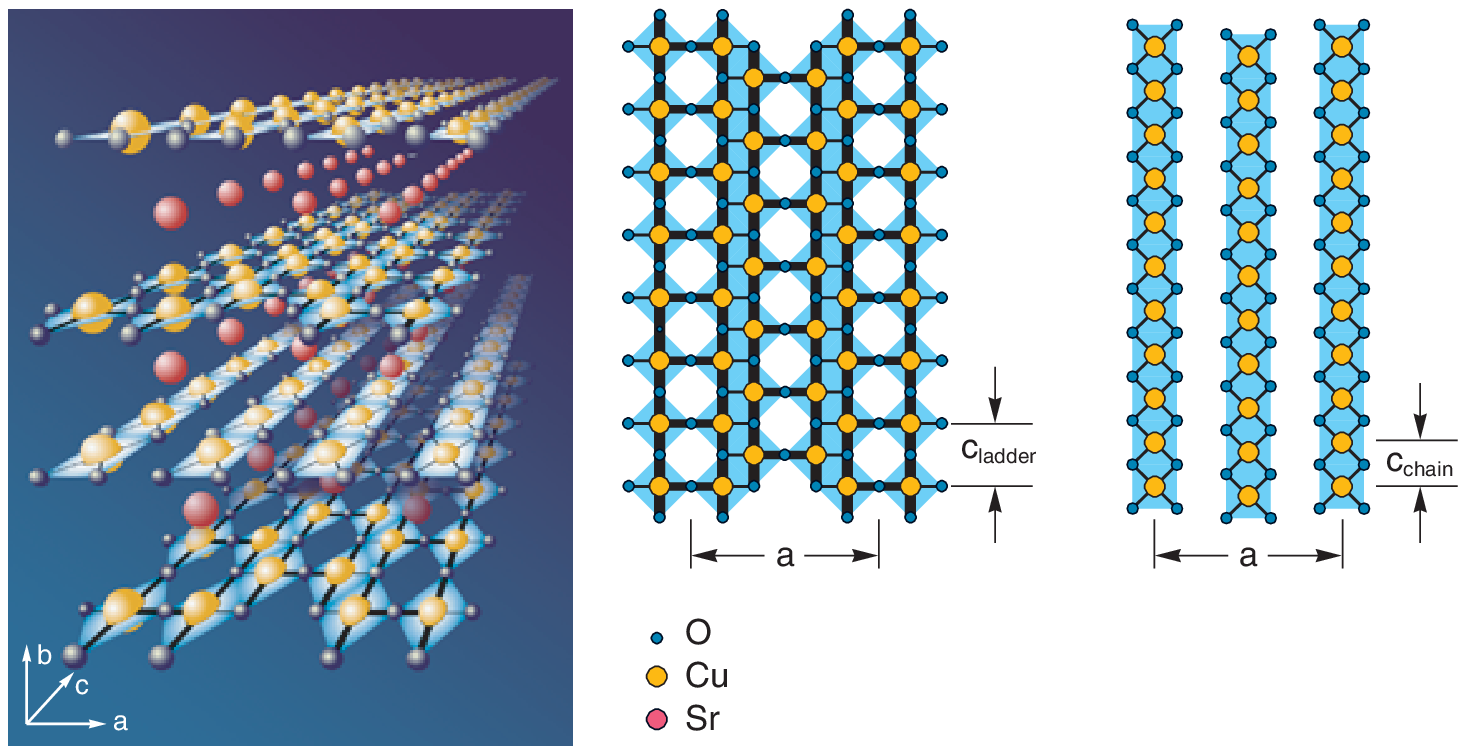,width=90mm}
}
\vspace{4mm}
\caption{\baselineskip24pt
A 3D sketch of the \SrCu crystal structure.
Three neighboring Cu$_{2}$O$_{3}$ ladder and CuO$_{2}$ chain subunits are 
shown separately. See \cite{McCarron,Siegrist} for deteils. 
}
\label{Structure}
\end{figure}

\SrCu is an experimental realization of a two-leg ladder structure 
(Fig.~\ref{Structure}).
The planes of the weakly coupled Cu$_{2}$O$_{3}$ ladders are stacked
along the crystallographic $b$-axis alternating with 1D CuO$_{2}$
edge-sharing chain sheets \cite{McCarron,Siegrist}.
The Cu-Cu distances in these two subunits are incommensurate but
satisfy an approximate relation $10 c_{chain} \simeq 7 c_{ladder}$.
The legs and rungs of the ladders are along the $c$- and
$a$-axes, respectively.
Because the average valence of Cu is +2.25, the compound is intrinsically 
doped with holes believed to reside mainly in the chain substructure; 
optical studies have estimated 0.07 holes per ladder Cu site
\cite{Osafune97}.

The self-doped compound \SrCu is an insulator with an Arrhenius 
temperature dependence 
of the dc resistivity \cite{McElfresh89}.
Microwave \cite{Kitano01} and magnetic resonance \cite{Takigawa88} 
studies have suggested the possibility of a charge ordering in the ladder 
planes that leads to insulating behavior. 
For ladders with higher hole concentrations, $d$-wave
like superconductivity was predicted to win the competition with
the charge-ordered state \cite{Rice96,Dagotto92}.
Superconductivity at 12~K was discovered in doped two-leg ladders by
applying both external and chemical pressure \cite{Uehara96}. 

\begin{figure}[ht]
    \baselineskip24pt
\centerline{
\epsfig{figure=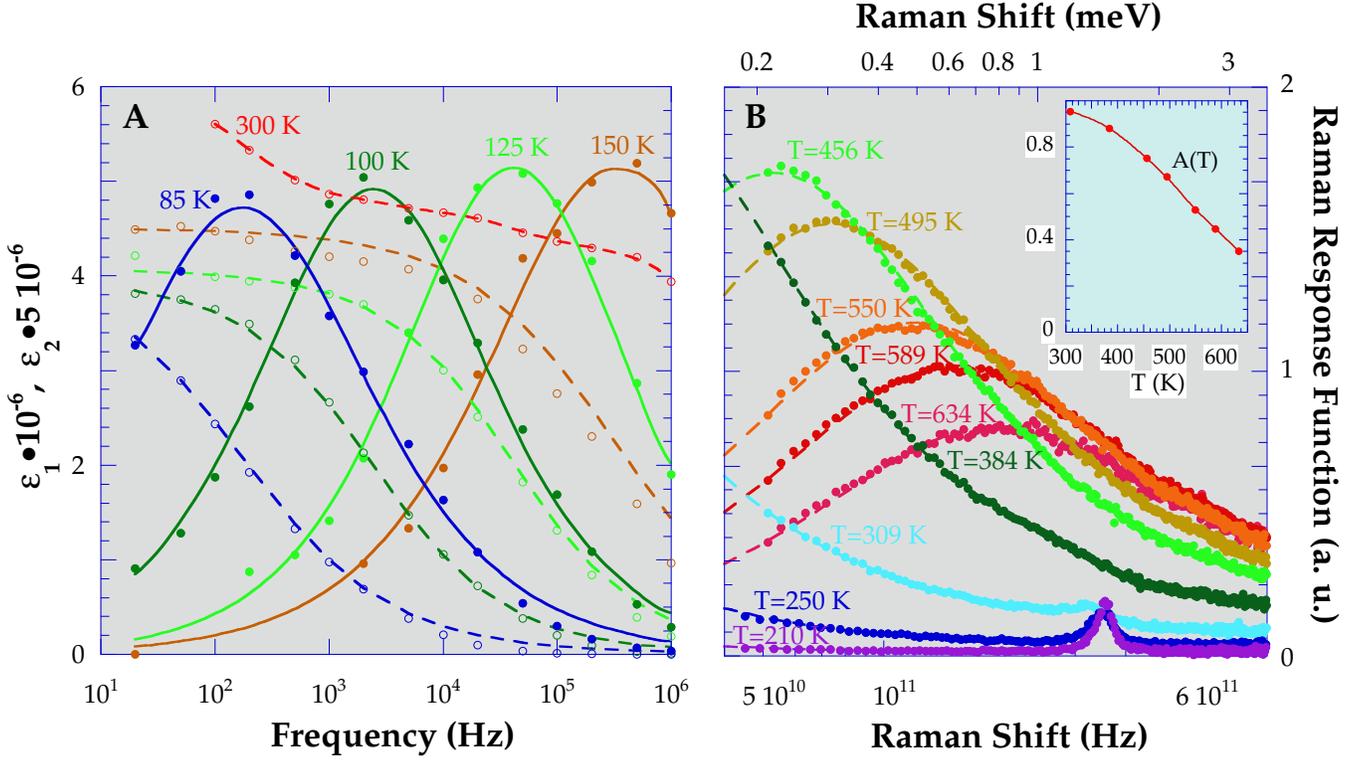,width=180mm}
}
\vspace{4mm}
\caption{\baselineskip24pt
(A) The temperature dependence of the measured real (open circles) and 
imaginary (solid circles) parts of the complex dielectric function between 
85 and 300~K.
The dashed and solid lines are guides for the eye.
(B) The temperature dependence of the Raman response function 
for the polarization of the incident and scattered photons parallel to the 
legs of the ladders: 
the data (circles) and the fit to eq. (\ref{QEP}) (dashed lines).
The spectra were acquired in a backscattering geometry using the 
7993~\AA \ excitation. 
The resonance at about 356~GHz at low temperatures is a phonon.
The inset shows the diminishing quasi-elastic scattering intensity, $A(T)$, 
with heating. 
}
\label{EpsilonRaman}
\end{figure}

\begin{figure}[ht]
\centerline{
\epsfig{figure=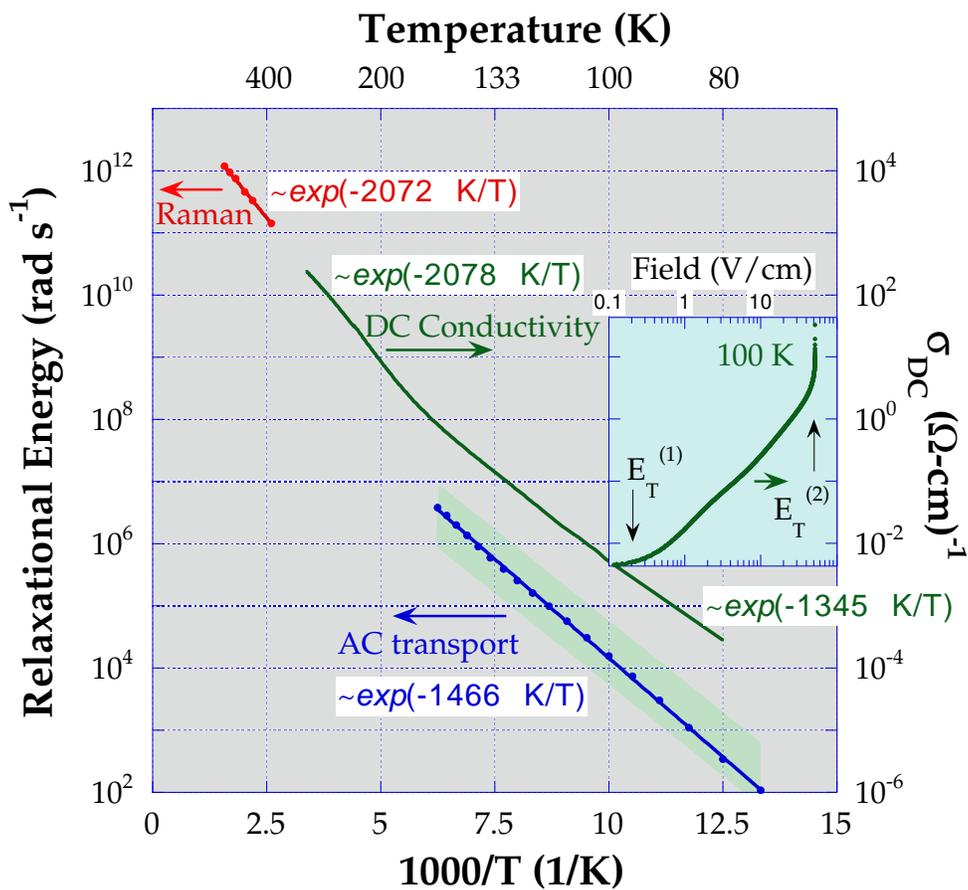,width=130mm}
}
\vspace{4mm}
\caption{\baselineskip24pt
The temperature dependence of the scattering rates (left scale) $\Gamma(T) =
\tau^{-1}$ from scaling of the ac dielectric (Fig.~\ref{Scaling}) and
Raman (Fig.~\ref{EpsilonRaman}B) response.
The scattering rate follows the activated behavior of the 
dc conductivity (right scale) over 10 decades of frequencies.
The inset shows strong nonlinearity in the dc conductivity as a function 
of electric field measured at 100~K.}
\label{Rates}
\end{figure}

Single crystals of \SrCu were grown as described in
\cite{Osafune97,Motoyama}. 
We have studied the low-frequency charge response by  
ac transport and Raman scattering experiments.
The temperature dependence of the $c$-axis complex dielectric function 
between $10$ and $10^{6}$~Hz (Fig.~\ref{EpsilonRaman}A) was measured with 
a capacitive technique.
The dc conductivity was measured by an electrometer (Fig.~\ref{Rates}).
Raman measurements (Fig.~\ref{EpsilonRaman}B) were performed from the 
(010) surface of the crystal as described in \cite{Gozar02}. 
We also report strong nonlinearity in the $c$-axis conductivity 
as a function of electric field.  

The temperature dependence of the low-frequency
Raman response function (Fig.~\ref{EpsilonRaman}B)
shows a broad overdamped excitation below 1~meV ($\sim250$~GHz).
This quasi-elastic scattering peak (QEP) rapidly shifts to lower
frequencies with cooling.
The data are fit by a relaxational form of the Raman response function
\begin{equation}
\chi'' (\omega, T) = A(T) \frac{\omega \Gamma(T)}{\omega^{2} +
\Gamma(T)^{2}}.
\label{QEP}
\end{equation}
The fit reveals a decrease in the QEP intensity, $A(T)$, with heating and
an Arrhenius temperature dependence for the scattering rate $\Gamma(T)$  
with an activation gap $\Delta \simeq 2072$~K. 
The dc conductivity exhibits activated behavior with a break in the
activation energy around $T^{*} \simeq 150$~K: $\Delta \simeq 1345$~K for
$T < T^{*}$ and $\Delta = 2078$~K above $T^{*}$, similar to the activation
energy observed for the Raman response function scattering rate 
$\Gamma(T)$ (Fig.~\ref{Rates}).
Similar quasi-elastic excitation has also been observed at high 
temperatures for superconducting Ca doped 
\SrCax ($0 < x < 12$) compounds with a higher concentration of holes in 
the ladder plane \cite{Gozar02}.

The frequency dependence of complex dielectric constant
$\epsilon = \epsilon_{1} + i \epsilon_{2}$ is shown in 
Fig.~\ref{EpsilonRaman}A for temperatures between 85 and 150~K.
The low-frequency response of $\epsilon_{2}$
shows overdamped, inhomogeneously
broadened peaks at characteristic frequencies $\nu_{0}(T)$ determined by
damping parameters $\Gamma(T) = 2 \pi \nu_{0}(T) = \tau^{-1}$
(where $\tau (T)$ is the relaxation time).
Strong relaxational peaks lead to a giant real part of the dielectric
response $\epsilon_{1}$ below $\nu_{0}(T)$ observed up to room temperature
and even above.
Such behavior is incompatible with any single-particle theory as it would 
imply energy gaps more than six orders of magnitude smaller than the 
thermal energy and suggests  
that the low frequency charge dynamics is driven 
by correlated collective behavior.

\begin{figure}[ht]
    \baselineskip24pt
\centerline{
\epsfig{figure=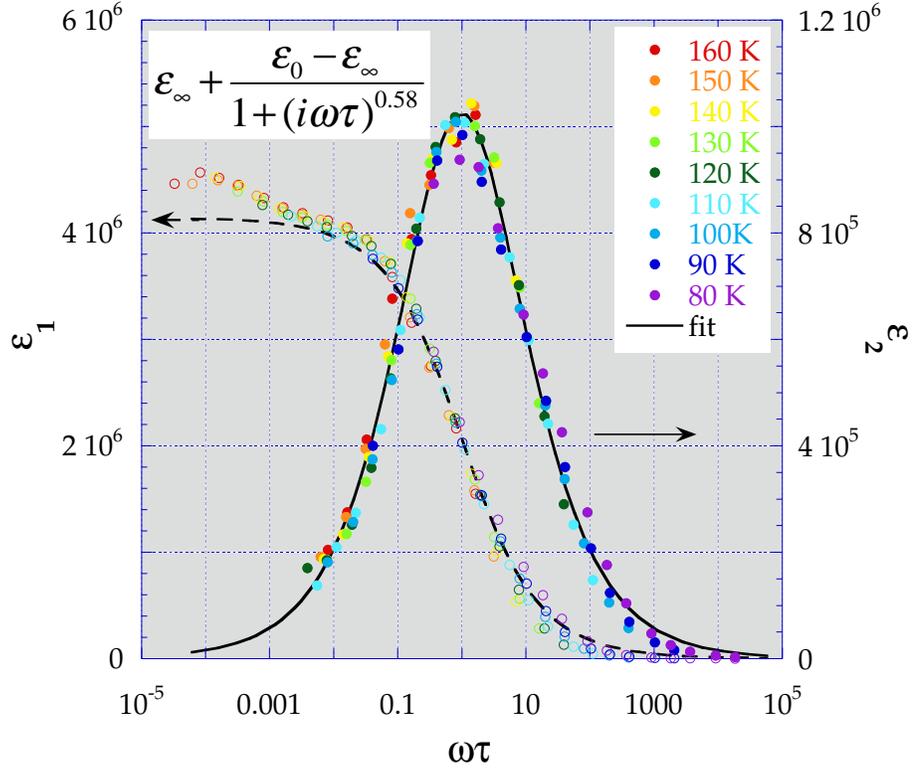,width=130mm}
}
\vspace{4mm}
\caption{\baselineskip24pt
Scaling, measured between 80 and 160~K in the 20~Hz~--~1~MHz frequency 
window, of the complex dielectric function (imaginary part, solid 
circles; real part, open circles) on the generalized Debye relaxational 
curve (\ref{Debye}). 
The obtained temperature dependent scattering rate $\Gamma(T) =
\tau^{-1}$ is shown in Fig.~\ref{Rates}.
}
\label{Scaling}
\end{figure}

In Fig.~\ref{Scaling} we scale all the measured complex dielectric
functions between 80 and 160~K on one universal generalized Debye
relaxational curve, 
\begin{equation}
\epsilon (\omega) = \epsilon_{\infty} + \frac{\epsilon_{0} -
\epsilon_{\infty}}{1 + [i \omega \tau(T)]^{1-\alpha}}
\label{Debye}
\end{equation}
where $\epsilon_{0}$ and $\epsilon_{\infty}$ are low- and
high-frequency dielectric constants and the fitted value $\alpha = 0.42$
characterizes the width of the distribution of relaxational times
\cite{Havriliak66}.
The temperature dependence of the scattering rates $\Gamma(T) =
\tau^{-1}$ from ac transport and Raman measurements are compared to the 
dc conductivity in Fig.~\ref{Rates}: 
The scattering rates $\Gamma(T)$ follow the activated behavior of dc 
conductivity over 10 decades of frequency.

We now discuss the implications of these observations.
First we note that recent microwave experiments by Kitano {\em et
al.} have reported a relatively small and narrow
peak between 30 and 100~GHz in the $c$-axis conductivity observed up to
moderately high temperatures \cite{Kitano01}.
This resonance is attributed by the authors to a collective
excitation of a pinned CDW.

In the charge or spin density state the condensed electrons do not
participate in the conduction process for small dc electric
fields.
The reason for this is the interaction between the condensate and
impurities and other lattice irregularities.
This pinning shifts the oscillator strength associated with the
collective modes to finite frequencies, with no contribution to
the dc conductivity that shows an activated behavior and is entirely
determined by quasi-particle excitations out of the condensate.
The Arrhenius law for the decay constant $\Gamma (T)$ suggests a
hydrodynamic origin for the Raman and low-frequency conductivity modes,
for which there is considerable precedent in studies of the dynamics of
pinned CDW and SDW systems
\cite{Gruner80,Wu84,Cava84,Cava86,Degiorgi91,GrunerBook}, 
and we compare the \SrCu
ladder system to well-established models of CDW
dynamics \cite{Littlewood87}.
The pinned collective mode can be described by an oscillator making a
collective contribution to the ac conductivity
       \begin{equation}
       \sigma_{coll} (\omega ) = \frac{1}{4 \pi} \frac{ - i \omega
       \Omega_p^2}{\Omega_o^2 -\omega^2 -i \gamma_o \omega}, 
       \label{scoll}
       \end{equation}
where $\Omega_p^2 = \rho_c^2/ m^*$ is the oscillator strength of the
mode with charge density $\rho_c$ and mass density $m^*$, $\Omega_0$ is the
pinning frequency, and $\gamma_0$ is an intrinsic damping coefficient.
To account for the longitudinal response (relevant for Raman scattering),
we must allow for screening of the collective mode. 
Then the longitudinal response function is
       \begin{equation}
       \epsilon_L (\omega) = \frac{ \Omega_p^2 } { \Omega_o^2 -\omega^2
       -i \gamma_o \omega - i \omega \Omega_p^2/(4 \pi \sigma_{qp} - i \omega
       \epsilon_\infty)}, 
       \label{long}
       \end{equation}
where $\sigma_{qp}$ is the dc "background" conductivity due
to quasiparticle transport.
Note that at low frequency, Eq.~(\ref{long}) reduces to a relaxational mode,
with amplitude $\Omega_p^2 / \Omega_0^2$ and decay constant
\begin{equation}
\Gamma(T) = 4 \pi \sigma_{qp} \frac {\Omega_0^2} {\Omega_p^2} = \frac
{4 \pi \sigma_{qp}} {\epsilon_0 - \epsilon_\infty}.
\label{Gamma}
\end{equation}
The proportionality of $\Gamma(T)$ and $\sigma_{qp}$ arises because the
collective charge oscillations are screened by thermally excited and
uncondensed quasiparticle carriers on the relevant ladder.
Although the longitudinal mode we are discussing is in principle neutral,
the effects of disorder in a strongly anisotropic system make it visible ---
in fact typically dominant ---
in the transverse response, and therefore it will contribute to ac
conductivity.
Such was the case in CDW systems
\cite{Gruner80,Wu84,Cava84,Cava86,Degiorgi91,GrunerBook}. 
The transfer of spectral weight to the longitudinal channel explains why 
the oscillator strength in the microwave region \cite{Kitano01} is 
anomalously small.

The plasma edge energy for \SrCu, determined from IR spectroscopy as the
peak position of the loss function $\Im [-1/\epsilon (\omega)]$, gives 
us an estimate
for $\Omega_p \simeq 3300$~\cm-1 \cite{Osafune97,MotoyamaOptics}.
Using the characteristic pinning frequency $\Omega_0 \simeq 30 -
100$~GHz suggested by
microwave measurements \cite{Kitano01}, we estimate the value of 
the low-frequency dielectric constant $\epsilon_0 \simeq 
\Omega_p^2/\Omega_0^2$ to be 
between $10^{6}$ and $10^{7}$, which is consistent with our direct
measurements.
Following expression (\ref{Gamma}) and the estimated values for 
$\epsilon_0$, we calculate the theoretical values for $\Gamma(T)$ using
the measured dc conductivity.
These calculated values are shown by the shaded green area in 
Fig.~\ref{Rates}.
They are in very good agreement with the relaxational energies extracted
from the peaks in $\epsilon_{2} (\omega)$ (see Fig.~\ref{Scaling}).
The overall consistency among the measured temperature dependences of the
dielectric response $\epsilon(\omega)$, the relaxation rate
$\Gamma(T)$, and the dc conductivity $\sigma_{qp}$ demonstrates the
applicability of the hydrodynamic model description for the low-frequency
collective charge dynamics in this system.    Moreover the
broad inhomogeneous response has exactly the character expected for
random pinning by disorder of an incommensurate density wave.

The small frequency $\Omega_0$, which describes the oscillation of the
collective mode about the pinned equilibrium position, implies a
weak restoring force.
Therefore, it is anticipated that a small dc electric field
can induce sliding density wave transport by translational motion of
the condensate.
In the inset of Fig.~\ref{Rates} we show conductivity at 100~K measured 
as a function of applied electric field.
Below a threshold field $E_{T}^{(1)} \simeq 0.2$~V/cm, the conductivity
obeys Ohm's law with the Arrhenius temperature dependence discussed
above.
However, for electric fields above the $E_{T}^{(1)}$ threshold, the
$I-V$ characteristics change from linear to
approximately quadratic, indicating an onset for sliding density
wave conductivity \cite{Zettl82}.
In this regime of relatively slow CDW sliding the predominant damping 
mechanism is the screening of internal electric fields produced by
local CDW deformations by backflow currents of uncondensed
quasi-particles \cite{Littlewood87}.

At much higher fields, above 50~V/cm, we observe a second threshold, 
$E_T^{(2)}$, with a very sharp rise of the current up to the limit of 
the current generator.
The calculated differential conductivity in this regime is enormous, more
than $10^{5}\ \Omega^{-1}$\cm-1, an estimate limited by contact effects
and most likely carried by inhomogeneous filamentary conduction.
We attribute this phenomenon to the crossover to a regime of free sliding CDW
first proposed by Fr\"{o}hlich as CDW superconductivity \cite{Frohlich}.
Such a crossover has been observed in conventional CDWs, and is expected
to occur once the density wave reaches velocities at which the
background quasiparticles can no longer screen the
response\cite{Littlewood88}. 
An upper limit of the field at this second threshold may be estimated from 
the average pinning frequency 
$E_T^{(2)} \leq (\Omega_0^2/\Omega_p^2) (\rho_c/Q)$,
where $\rho_c$ is the collective charge density and $2\pi/Q$ is the period
of the density wave.
Assuming $2e$ per ladder contributing to the density wave, we obtain
$E_T^{(2)} \leq 100\; V/cm$, again consistent with our measurements.
Note that the true threshold for deformable sliding of the density wave
is always much lower.

Now we turn to high temperature Raman measurements.
The Raman response function is proportional to $\Im {\epsilon_L}$.
The comparison of the Raman and ac transport data is not straightforward
because only in the Raman data above room temperature $\Gamma(T)$ is 
large enough for a peak to be observed.
Although $\Gamma(T)$ extracted from the Raman data exhibits activated behavior
with a gap consistent with dc conductivity above $T^{*}$, the
values for $\Gamma(T)$ are about 50 times the relaxational
energies predicted from eq.~(\ref{Gamma}).
However, the lowering of the peak intensity with increasing
temperature (see Fig.~\ref{EpsilonRaman}B inset) suggests that there is 
a reduction of the density wave amplitude $\rho_c$ which would produce 
a concomitant increase in $\Gamma$.
Further enhancement in the scattering rate may come from additional
relaxation due to the presence of low lying states seen at temperatures
above $T^{*}$ by magnetic resonance \cite{Takigawa88}, high frequency Raman
scattering \cite{Gozar01} and $c$-axis optical conductivity
\cite{MotoyamaOptics}.

All our results  have clear quantitative parallels with sliding density
wave transport  phenomena observed in established C/SDW materials,
yet there must be a number of important microscopic differences from
conventional weak-amplitude charge- and spin-density waves.
The C/SDW correlation in \SrCu is a high-temperature phenomenon 
that we observe (with diminishing amplitude) up to the highest measured
temperature, 630~K. 
Such high-temperature correlations can not be supported by phonons, 
which suggests that the charge/spin correlations arise from strong spin 
exchange interactions with characteristic energy scale  $J \simeq 
1300$~K \cite{Gozar01}.
The magnetic excitations in spin $\frac{1}{2}$ two-leg
ladders are resonating valence bond (RVB) quantum states
\cite{Rice96,Sachdev,Gozar01,Anderson87}, different from the gapless
classical spin waves in the SDW systems.
Theoretical calculations for a doped two-leg spin ladder suggest
that in an RVB environment the holes are paired in a state of
approximate $d$-wave symmetry with a few lattice spacings in size
\cite{Sigrist94,Troyer96,White97}.
The superconducting condensation of bound pairs is competing with
a crystalline order of these pairs in a CDW state \cite{Dagotto92}.
If the fundamental current-carrying object in the doped singlet ladders
is a bound pair of holes, the hydrodynamic mode we have discussed involves 
local displacement of ``crystallized'' pairs, screened by charged 
excitations across the CDW gap, which would account for the appearance 
of the conductivity as the relevant parameter to describe backflow.
Nevertheless, our observation of a weakly pinned mode --- which implies 
a large 
stiffness for charged fluctuations of the order parameter --- is surprising 
in the context of models with strong short-range correlations, where one 
would expect that lattice pinning effects are strong.

\newpage

\newpage

\newpage

\end{document}